\documentclass[prb,twocolumn,showpacs]{revtex4}%
\usepackage{amsfonts}
\usepackage{amsmath}
\usepackage{amssymb}
\usepackage{graphicx}%
\begin{document}
\title{Current noise of a resonant tunnel junction coupled to a nanomechanical oscillator}
\author{M. Tahir}
\altaffiliation[Permanent address: ]{Department of Physics, University of Sargodha, Sargodha, Pakistan}
\email[E-Mail: ]{m.tahir@uos.edu.pk}
\thanks{M.Tahir would like to acknowledge the support of the Pakistan Higher Education Commission (HEC).}
\author{A. MacKinnon}
\email[E-Mail: ]{a.mackinnon@imperial.ac.uk}
\affiliation{The Blackett Laboratory, Imperial College London, South
Kensington campus, London SW7 2AZ, United Kingdom}
\begin{abstract}
We present a theoretical study of current noise of a resonant tunnel junction
coupled to a nanomechanical oscillator within the non-equilibrium Green's
function technique. An arbitrary voltage is applied to the tunnel junction and
electrons in the leads are considered to be at zero temperature. The
properties of the phonon distribution of the nanomechanical oscillator
strongly coupled to the electrons on the dot are investigated using a
non-perturbative approach. An analytical calculations and numerical results
for the current-voltage, shot noise and the corresponding Fano factor as a
function of applied bias show significant features of the nanomechanical
oscillator coupling dynamics. This will provide useful insight for the design
of experiments aimed at studying the quantum behavior of an oscillator.

\end{abstract}
\startpage{01}
\endpage{02}
\maketitle

\section{Introduction}

In recent years, there has been a great interest in the measurement of current
noise near the quantum limit of displacement sensitivity using single electron
transistors (SET) and nanomechanical oscillators\cite{1,2,3,4,5,6}. Current
noise is the non-equilibrium fluctuation which is caused by the discreteness
of charged carriers. When the size of an electromechanical system reaches the
nanometer scale, the current noise problem becomes a very interesting aspect
of the NEMS\cite{7,8,9,10}\ based devices. In NEMS\ devices it is a
fundamental physical signature, which can provide us additional information
regarding quantum transport, in addition to differential conductance and
current voltage characteristics. The electrons traveling through the device
become correlated in the same channel and the same probe, as well as in
different channels and different probes. For a mesoscopic system, the
electrons are correlated due to coherent transport and they are governed by
the Fermi distribution and Coulomb blockade. Noise is caused by the randomness
of electron scattering and it has been extensively studied in different types
of mesoscopic structures. For uncorrelated electrons travelling through a
macroscopic conductor, shot noise $S(0)$ at zero frequency is given by the
Poisson value $S(0)=2eI$, where I is the net average current flowing through
the system. However, for a mesoscopic system, electrical current correlations
are dominated by the Coulomb blockade and a Fermi distribution can lead to the
deviation of shot noise from a Poissonian (F=1) form. An important parameter
for describing this type of shot noise is the Fano factor F$=S(0)/2eI$, using
which the shot noise can be classified as sub-Poissonian (F
$<$
1) or super-Poissonian (F
$>$
1). Similarly, phonon assisted features have been observed in molecular
systems\cite{11,12,13,14} with strong interactions, which is beyond the scope
of perturbation theory.

In general, there are two different theoretical formulations that can be used
to study the quantum transport in nanoscopic systems under voltage bias.
Firstly, a generalized quantum master equation
approach\cite{15,16,17,18,19,20} and secondly, the non-equilibrium Green's
function formula\-tion\cite{21,22,23,24,25}. The former leads to a simple rate
equation, where the coupling between the dot and the leads is considered as a
weak perturbation and the electron-phonon interaction is also considered very
weak. In the latter case one can consider strong leads to system and
electron-phonon coupling. The non-equilibrium Green's function technique is
able to deal with a very broad variety of physical situations related to
quantum transport at molecular\cite{26,27,28} levels. It can deal with strong
non-equilibrium situations and very small to very large applied bias. In most
of the theoretical work on NEMS devices, the mechanical degree of freedom has
been described classically/semiclassically\cite{15,16} using the master
equation approach or quantum mechanically\cite{17,18,19,20,34,35} using the
perturbation approach for the electron-phonon interactions. Furthermore, Shot
noise for electron transport in molecular devices was investigated within a
scattering theory approach\cite{29,30}. Phonon effects in the noise spectrum
were studied recently\cite{31,32,33}, in connection with NEMS and both single
electronic levels in quantum dots and single molecules. In Armour's
work\cite{15}, the mechanical part was also treated classically, including the
damped oscillator and assuming a weak electron tunneling process. This
approach is based on a perturbation, weak coupling and large applied bias
approximations, whereas the Keldysh non-equilibrium Green's function
formulation can treat the system-leads and electron-phonon coupling with
strong interactions for both small and large applied bias voltage. Moreover,
these theories fail to explain the low bias regime. The transport properties
have been described and discussed semi-classically/classically but need a
complete quantum mechanical description. A theory beyond these cases is
required to further refine experiments to investigate the quantum transport
properties of NEMS\ devices.

In the present work, we employ the non-equilibrium Green's function method to
discuss the current-voltage and shot noise properties of a NEMS device. This is
a fully quantum mechanical formulation whose basic approximations are very
transparent, as the technique has already been used to study transport
properties in a wide range of systems. The main differences between existing
work and ours are: in most of the existing literature a very large chemical
potential difference is considered while we are able to include a range from
very small to very large. In our calculation the inclusion of the oscillator
is not perturbative as the STS experiments\cite{11,12,13,14} are beyond the
range of perturbation theory. Hence, an approach is required beyond the
quantum master equation or linear response. Hence, our work provides an exact
analytical solution to the current--voltage, shot noise, coupling of leads
with the system, very small to very large chemical potential difference and
includes both the right and left Fermi level response regimes. For simplicity,
we used the wide--band approximation, where the coupling between the leads and
the dot is taken to be independent of energy. This provides a way to perform
transient transport calculations from first principles while retaining the
essential physics of the electronic structure of the dot and the leads.
Another advantage of this method is that it treats the infinitely extended
reservoirs in an exact way in the present system, which may give a better
understanding of the essential features of NEMS in a more appropriate quantum
mechanical picture.

\section{Model calculations}

We consider a single quantum dot connected to two identical metallic leads via
tunnel junctions (SET device)\cite{15,19,35}. A single nanomechanical
oscillator is coupled to the electrons on the dot and the applied gate voltage
is used to tune the single level of the dot. This model represents a close
analog of an electromechanical displacement detector and contains the
essential features of the recent experiments performed with a
resonator\cite{1,2}, or a vibrating beam of crystal or cantilever coupled to
the SET device\cite{3,4,5,6}. In most of these experiments, the motion of a
resonator (nanomechanical oscillator) may be detected by capacitively coupled
electrodes placed on the resonator and biasing the electrode at a constant
voltage. The dot has a small capacitance due to its small diameter (nm) and
thus has charging energy. This charging energy exceeds the thermal energy in
these experiments. For this reason we consider that only one excess electron
may occupy the device. To ensure this condition we are working at zero temperature.

In the present simple model, both the electronic and the mechanical degrees of
freedom have been treated quantum mechanically, where the nanomechanical
oscillator is represented in terms of creation and annihilation operators. We
neglect the spin degree of freedom and electron-electron interaction effects
and consider the simplest possible model system. We also neglect the effects
of finite electron temperature of the lead reservoirs and damping of the
oscillator. Our model consists of the individual entities such as the single
quantum dot with perpendicular oscillator on it and the left and right leads
in their ground states at zero temperature. The Hamiltonian of our simple
system\cite{24,25,26,28} is%

\begin{eqnarray}
H_{0} &=& H_{ph-dot}+H_{leads} \label{1}\\%
H_{\mbox{\scriptsize dot-ph}} &=& \left[  \epsilon_{0}+\eta(b^{\dagger}+b)\right]
c_{0}^{\dag}c_{0}+H_{ph}\,,\label{2}\\%
H_{ph} &=& \frac{\hat{p}^{2}}{2\mu}+\frac{1}{2}\mu\omega
^{2}\hat{x}^{2}=\hslash\omega(b^{\dagger}%
b+{\mathchoice{{\textstyle{\frac12}}}{{\textstyle{\frac12}}}{{\scriptstyle{1/2}}}{{\scriptscriptstyle{1/2}}}}%
)\,,\label{3}%
\end{eqnarray}
where $\epsilon_{0}$ is the single energy level of electrons on the dot with
$c_{0}^{\dag},c_{0}$ the corresponding creation and annihilation operators.
The parameter $\eta=\lambda l/\sqrt{2}$ physically represents an effective
electric field in the capacitor formed by the oscillator and the dot electrons
, which we shall call coupling strength between the oscillator and the
electrons on the dot given as $\lambda l=eEl$, where $e$ is the charge of
electron, $E$ is the strength of the electric field and $l=\sqrt{\frac
{\hslash}{\mu\omega}}$ is the zero point amplitude of the oscillator with mass
$\mu$. Here we assume that the energy of the resonant level depends linearly
on the oscillator coordinate. The frequency of the nanomechanical oscillator
is $\omega$ and $b^{\dagger}$, $b$ are the raising and lowering operator of
the phonons given as%

\begin{equation}
\hat{x}=\frac{l}{\sqrt{2}}(b^{\dagger}+b), \label{4}%
\end{equation}
and%

\begin{equation}
\hat{p}=\frac{i\mu\omega}{\sqrt{2}}l(b^{\dagger}-b),
\label{5}%
\end{equation}
The remaining elements of the Hamiltonian are%

\begin{eqnarray}
H_{\mbox{\scriptsize leads}} &=& \sum_{j}\epsilon_{j}c_{j}^{\dagger}c_{j},
\label{6}\\%
\Delta H_{\alpha}=H_{\mbox{\scriptsize leads-dot}}
&=&\frac{1}{\sqrt{N}}\sum
_{j}V_{\alpha}\left(  c_{j}^{\dagger}c_{0}+c_{0}^{\dagger}c_{j}\right)  ,
\label{7}%
\end{eqnarray}
where $N$ is the total number of states in the lead, $V_{\alpha}$ is the
hopping between the dot and the leads $\alpha=L,R$, $j$ represents the
channels in one of the leads. For the second lead the Hamiltonian can be
written in the same way.

The total Hamiltonian of the system is thus $H=H_{0}+\Delta H_{\alpha}\,$. We
write the eigenvalues and the eigenfunctions of $H_{\mbox{\scriptsize
dot-ph}}$ as
\begin{eqnarray}
\epsilon &=& \epsilon_{0}+\hslash\omega
(n+{\mathchoice{{\textstyle{\frac12}}}{{\textstyle{\frac12}}}{{\scriptstyle{1/2}}}{{\scriptscriptstyle{1/2}}}}%
)-\Delta\label{8}\\%
\Psi_{n}(K,x_{0}=0) &=& A_{n}\exp[-{\textstyle\frac{l^{2}K^{2}}{2}}]H_{n}(lK)\,,
\label{9}\\%
\Psi_{n}(K,x_{0}\neq0) &=& \Psi_{n}(K,x_{0}=0)\exp[-{\mathrm{i}}Kx_{0}] \label{10}%
\end{eqnarray}
for the occupied, $x_{0}\not =0$ and unoccupied, $x_{0}=0$, dot respectively,
where $A_{n}={1}/{\sqrt{\sqrt{\pi}2^{n}n!l}}$,   $\Delta={\lambda^{2}}/{\mu\omega^{2}}$. 
$x_{0}={\lambda}/{\mu\omega^{2}}$ is the shift of the oscillator due to the
coupling to the electrons on the dot and $H_{n}(lK)$ are the usual Hermite
polynomials. Here we have used the fact that the harmonic oscillator
eigenfunctions have the same form in both real and Fourier space.

In order to transform between the representations for the occupied and
unoccupied dot we require the matrix with elements%

\begin{equation}
\Phi_{n,m}=\int\Psi_{n}^{\ast}(K,x_{0}=0)\Psi_{m}(K,x_{0}\neq0)\,{\mathrm{d}%
}K, \label{11}%
\end{equation}
which may be simplified\cite{36} as%
\begin{equation}
\Phi_{n,m}  = \sqrt{\frac{2^{m-n}n!}{m!}}\exp\left(  -{\textstyle\frac{1}%
{4}}x^{2}\right)  \left(  {\textstyle\frac{1}{2}}\mathrm{i}x\right)
^{m-n}L_{n}^{m-n}\left(  {\textstyle\frac{1}{2}}{x^{2}}\right) \label{12}
\end{equation}
for $n\leq m$, where $x=\frac{x_{0}}{l}$ and $L_{n}^{m-n}(x)$ are the
associated Laguerre polynomials. Note that the integrand is symmetric in $m$
and $n$ but the integral is only valid for $n\leq m$. Clearly the result for
$n>m$ is obtained by exchanging $m$ and $n$ in equation~(\ref{12}) to obtain
\begin{eqnarray}
\Phi_{n,m}&=&\sqrt{\frac{2^{|m-n|}\min[n,m]!}{\max[n,m]!}}
\exp\left(-\textstyle\frac{1}{4}x^{2}\right)\nonumber\\  
&&\times\left(  \textstyle\frac{1}{2}%
\mathrm{i}x\right)  ^{|m-n|}L_{\min[n,m]}^{|m-n|}\left(  \textstyle\frac{1}%
{2}{x^{2}}\right)  \,. \label{13}%
\end{eqnarray}
The position of the resonant level with respect to the chemical potential in
the leads is thus affected by the displacement of the nanomechanical
oscillator, which in turn affects the transport properties of the junction
through the device.

The particle current $I_{\alpha}$ into the interacting region from the lead is
related to the expectation value of the current operator $I_{\alpha}%
=\mathop{\mathrm{Tr}}(\rho\hat{J}_{\alpha})$\cite{37,38,39,40}, where%
\begin{equation}
\hat{J}_{\alpha}=\frac{ie}{\hbar}\sum_{j}\{V_{0,\alpha}c_{0}^{\dagger
}c_{\alpha j}-c_{\alpha j}^{\dagger}c_{0}V_{0,\alpha}\}\,, \label{14}
\end{equation}
$\rho=-iG^{<}$ and $\rho$\ is the density matrix written in terms of
the lesser Green's function. Eq.~(\ref{14}) can be written in terms of lesser
Green's function as%

\begin{equation}
I_{\alpha}(t)=\frac{e}{\hbar}\{G_{0,\alpha}^{<}(t,t)V_{\alpha,0}%
(t)-V_{0,\alpha}^{\ast}(t)G_{\alpha,0}^{<}(t,t)\}, \label{15}%
\end{equation}
where we have the following relations%

\begin{eqnarray}
G_{o,\alpha}^{<}(t,t)&=&%
{\displaystyle\int}
dt^{\prime}\{G_{0,0}^{r}(t,t^{\prime})V_{0,\alpha}(t^{\prime})g_{\alpha
,\alpha}^{<}(t^{\prime},t)\nonumber\\
&&+G_{0,0}^{<}(t,t^{\prime})V_{0,\alpha}(t^{\prime
})g_{\alpha,\alpha}^{a}(t^{\prime},t)\} \label{16}%
\end{eqnarray}
and%

\begin{eqnarray}
G_{\alpha,0}^{<}(t,t)&=&%
{\displaystyle\int}
dt^{\prime}\{g_{\alpha,\alpha}^{r}(t,t^{\prime})V_{\alpha,0}(t^{\prime
})G_{0,0}^{<}(t^{\prime},t)\nonumber\\
&&+g_{\alpha,\alpha}^{<}(t,t^{\prime})V_{\alpha
,0}(t^{\prime})G_{0,0}^{a}(t^{\prime},t)\}, \label{17}%
\end{eqnarray}
where $g_{\alpha,\alpha}^{r,(a),(<)}(t,t^{\prime})$ refers to the unperturbed
states of the leads.
\begin{widetext}
\begin{eqnarray}
I_{\alpha}(t)  &=& \frac{e}{\hbar}%
{\displaystyle\int}
\mathop{\mathrm{Tr}}\left\{(G_{0,0}^{r}(t,t^{\prime})V_{0,\alpha}(t^{\prime
})g_{\alpha,\alpha}^{<}(t^{\prime},t)
+G_{0,0}^{<}(t,t^{\prime})V_{0,\alpha
}(t^{\prime})g_{\alpha,\alpha}^{a}(t^{\prime},t))V_{\alpha,0}(t)\right.\nonumber\\
&&\left.-V_{0,\alpha}^{\ast}(g_{\alpha,\alpha}^{r}(t,t^{\prime})V_{\alpha
,0}(t^{\prime})G_{0,0}^{<}(t^{\prime},t)
+g_{\alpha,\alpha}^{<}(t,t^{\prime
})V_{\alpha,0}(t^{\prime})G_{0,0}^{a}(t^{\prime},t))\right\}dt^{\prime}
\,.\label{18}
\end{eqnarray}
Using the fact that 
\[
\Sigma_{0,0,\alpha}^{r,(a),(<)}(t^{\prime},t)=V_{0,\alpha
}^{\ast}(t^{\prime})g_{\alpha,\alpha}^{r,(a),(<)}(t^{\prime},t)V_{\alpha
,0}(t),
\]
we can simplify the above equation as%

\begin{equation}
I_{\alpha}(t)  = \frac{e}{\hbar}%
{\displaystyle\int}
\mathop{\mathrm{Tr}}\left\{G_{0,0}^{r}(t,t^{\prime})\Sigma_{0,0,\alpha}%
^{<}(t^{\prime},t)
+G_{0,0}^{<}(t,t^{\prime})\Sigma_{0,0,\alpha}^{a}(t^{\prime
},t)
-\Sigma_{0,0,\alpha}^{r}(t,t^{\prime})G_{0,0}^{<}(t^{\prime},t)
-\Sigma
_{0,0,\alpha}^{<}(t,t^{\prime})G_{0,0}^{a}(t^{\prime},t)]\right\}dt^{\prime}\,.\label{19}
\end{equation}
In the stationary regime all of the Green's functions depend on time order
($t-t^{\prime}$), yielding the Fourier/Laplace transform of equation~(\ref{19}) as%

\begin{equation}
I_{\alpha}  = \frac{e}{\hbar}%
{\displaystyle\int}
\frac{dE}{2\pi}\mathop{\mathrm{Tr}}\left\{G_{0,0}^{r}(E)\Sigma_{0,0,\alpha}^{<}(E)+G_{0,0}%
^{<}(E)\Sigma_{0,0,\alpha}^{a}(E)
 -\Sigma_{0,0,\alpha}^{r}(E)G_{0,0}^{<}(E)-\Sigma_{0,0,\alpha}^{<}%
(E)G_{0,0}^{a}(E)]\right\},\label{20}
\end{equation}
\end{widetext}

In order to calculate the analytical results and to discuss the numerical
quantum dynamics of the nanomechanical system, our focus is firstly to derive an
analytical relation for the retarded self-energy. The self-energy
represents the contribution to the dot energy, due to interactions between the
dot and the leads. In obtaining these results we use the
wide--band approximation where the retarded self--energy of the dot due to
each lead is considered to be energy independent and is given by%

\begin{equation}
\Sigma_{n_{0},n_{0},\alpha}^{r}(E)=\Delta H_{\alpha}^{\ast}g_{\alpha,\alpha
}^{r}\left(  E-(n_{0}%
+{\mathchoice{{\textstyle{\frac12}}}{{\textstyle{\frac12}}}{{\scriptstyle{1/2}}}{{\scriptscriptstyle{1/2}}}}%
)\hslash\omega\right)  \Delta H_{\alpha}, \label{21}%
\end{equation}
where off-diagonal element of matrix, $\Sigma_{n_{0},n_{0}^{\prime},\alpha
}^{r}(E)$, are zero and $g_{\alpha,\alpha}^{r}(E-(n_{0}+\frac{1}{2}%
)\hslash\omega)$\ is the uncoupled Green's function in the leads as%
\begin{eqnarray}
g_{\alpha,\alpha}^{r}(E-(n_{0}%
+{\mathchoice{{\textstyle{\frac12}}}{{\textstyle{\frac12}}}{{\scriptstyle{1/2}}}{{\scriptscriptstyle{1/2}}}}%
)\hslash\omega)
&=& \frac{1}{N}\sum_{j}g_{\alpha,j}^{r}(E-(n_{0}%
+{\mathchoice{{\textstyle{\frac12}}}{{\textstyle{\frac12}}}{{\scriptstyle{1/2}}}{{\scriptscriptstyle{1/2}}}}%
)\hslash\omega)\nonumber\\
&=& \frac{1}{N}\overset{+\infty}{\underset{-\infty}{%
{\displaystyle\int}
}}\frac{Nn_{\alpha}d\varepsilon_{\alpha}}{E-(n_{0}+\frac{1}{2})\hslash
\omega-\varepsilon_{\alpha}}\label{22}\nonumber\\
\end{eqnarray}
where $\underset{j}{%
{\displaystyle\sum}
}\mapsto\overset{+\infty}{\underset{-\infty}{%
{\displaystyle\int}
}}Nn_{\alpha}d\varepsilon_{\alpha},$ $j$ stands for every channel in each
lead and $n_{\alpha}$ is the density of states in lead $\alpha$. With the
help of equation~(\ref{22}), the retarded self-energy may be written as%
\begin{eqnarray}
\Sigma_{n_{0},n_{0},\alpha}^{r}(E)  &=& \Delta H_{0\alpha}\frac{1}{N}%
\overset{+\infty}{\underset{-\infty}{%
{\displaystyle\int}
}}\frac{Nn_{\alpha}d\varepsilon_{\alpha}}{E-(n_{0}+\frac{1}{2})\hslash
\omega-\varepsilon_{\alpha}}\Delta H_{\alpha 0}\nonumber\\
&=&-\left\vert V_{0,\alpha}\right\vert ^{2}n_{\alpha}\times(\pi
i)=\frac{-i\Gamma_{\alpha}}{2}\label{23}
\end{eqnarray}
which is now independent of $E$, $\Gamma_{\alpha}=2\pi\left\vert V_{0,\alpha
}\right\vert ^{2}n_{\alpha}$, $\alpha$\ representing the L or R leads and the
retarded self energy is now independent of the oscillator's index
($n_{0},n_{0})$. Hence, it can be written as $\Sigma_{n_{0}n_{0},\alpha}%
^{r}(E)=(\Sigma_{n_{0},n_{0},\alpha}^{a}(E))^{\ast}=-\frac{i\Gamma_{\alpha}%
}{2}$.

We solve Dyson's equation using H$_{dot-lead}$ as a perturbation. For the more
general systems we aim to treat in the future, this is a reasonable small
parameter. In the present case, however, we can find an exact solution. The
retarded and advanced Green's functions on the dot, with the phonon states in
the representation of the unoccupied dot, may be written as%

\begin{equation}
G_{nn_{0}}^{r(a)}(E)=\sum_{m}\Phi_{n,m}g_{m}^{r(a)}(E)\Phi_{n_{0},m}^{\ast}\,,
\label{24}%
\end{equation}
where $g_{m}^{r,(a)}(E)$ is the retarded (advanced) Green's function on the
occupied dot,
\begin{equation}
g_{m}^{r,(a)}(E)=\left[  E-\epsilon_{0}%
-(m+{\mathchoice{{\textstyle{\frac12}}}{{\textstyle{\frac12}}}{{\scriptstyle{1/2}}}{{\scriptscriptstyle{1/2}}}}%
)\hslash\omega+\Delta\pm\mathrm{i}\Gamma\right]^{-1}, \label{25}
\end{equation}
assuming that $\Gamma_{L}=\Gamma_{R}=\Gamma$.

The lesser Green's function in the presence of the nanomechanical oscillator
including the dot and the leads is written as%

\begin{equation}
G_{n,n^{\prime}}^{<}(E)=\sum_{n_{0},n_{0}^{\prime}}G_{n,n_{0}}^{r}%
(E)\Sigma_{n_{0}n_{0}^{\prime}}^{<}(E)G_{n_{0}^{\prime},n^{\prime}}^{a}(E),
\label{26}%
\end{equation}
with $\Sigma_{n_{0},n_{0}^{\prime}}^{<}(E)$\ being the the lesser self energy
which is given as%
\begin{equation}
\Sigma_{n_{0},n_{0}^{\prime}}^{<}(E)=\Sigma_{n_{0},n_{0}^{\prime},L}%
^{<}(E)+\Sigma_{n_{0},n_{0}^{\prime},R}^{<}\,(E), \label{27}
\end{equation}
where the off-diagonal element of matrix, $\Sigma_{n_{0},n_{0}^{\prime}%
,\alpha}^{<}$, are zero and the diagonal ($n_{0}^{\prime}=n_{0}$) element of
the lesser self--energy may be written as%
\begin{eqnarray}
\lefteqn{\Sigma_{n_{0},n_{0},\alpha}^{<}(E)}\nonumber\\
  &  =&\mathrm{i}\Gamma_{\alpha}\int
d\varepsilon_{\alpha}f_{\alpha}(\varepsilon_{\alpha})\,B_{n_{0}}%
\delta(E-\varepsilon_{\alpha}-(n_{0}%
+{\mathchoice{{\textstyle{\frac12}}}{{\textstyle{\frac12}}}{{\scriptstyle{1/2}}}{{\scriptscriptstyle{1/2}}}}%
)\hslash\omega)\nonumber\\
&=&\mathrm{i}\Gamma_{\alpha}f_{\alpha}(\varepsilon)\,B_{n_{0}},\label{28}
\end{eqnarray}
where $\varepsilon=E-n_{0}%
+({\mathchoice{{\textstyle{\frac12}}}{{\textstyle{\frac12}}}{{\scriptstyle{1/2}}}{{\scriptscriptstyle{1/2}}}}%
)\hslash\omega$, $f_{\alpha}(\varepsilon_{\alpha})$\ and $f_{\alpha
}(\varepsilon)$\ are the Fermi distribution functions of the left ($\alpha=L$)
and right ($\alpha=R$)\ leads, which have different chemical potentials under
a voltage bias and $B_{n_{0}}$\ is the Boltzmann factor for the oscillator
state. Index $n_{0}$ determines the statistical occupation probability of the
phonon state $\left\vert n_{0}\right\rangle $ at finite temperature and
therefore the accessibility of particular conduction channels is determined by
a weight factor of the Boltzmann distribution function.

Following equation~(\ref{20}), the formula for the current through each of the leads
is written in terms of oscillator indices as%

\begin{eqnarray}
I_{\alpha}&=&\frac{e}{2\pi\hbar}\sum_{n_{0},n}\int\biggl\{\Sigma_{n_{0},n,\alpha}%
^{<}(E)\left(G_{n,n_{0}}^{r}(E)-G_{n,n_{0}}^{a}(E)\right)\biggr.\nonumber\\
&&+\biggl.\left(\Sigma_{n_{0},n,\alpha}%
^{a}(E)-\Sigma_{n_{0},n,\alpha}^{r}(E)\right)G_{n,n_{0}}^{<}(E)\biggr\}dE \label{29}%
\end{eqnarray}
With the help of equation~(\ref{29}), the net current through the dot and the leads
with the oscillator on the dot is written as%
\begin{widetext}
\begin{eqnarray}
I  &  =&\frac{I_{L}-I_{R}}{2}\label{30}\\
& =& \frac{e}{4\pi\hbar}\sum_{n_{0},n}\int\biggl\{
\left(  \Sigma_{n_{0},n,L}^{<}(E)-\Sigma_{n_{0},n,R}^{<}(E)\right)\left(  G_{n,n_{0}}^{r}(E)-G_{n,n_{0}}^{a}(E)\right)
\biggr.\nonumber\\
&&\biggl.+\left[\left(  \Sigma_{n_{0},n,L}^{a}(E)-\Sigma_{n_{0},n,L}^{r}(E)\right)
-(\Sigma_{n_{0},n,R}^{a}(E)-\Sigma_{n_{0},n,R}^{r}(E))\right]G_{n,n_{0}}%
^{<}(E)
\biggr\}  \,\mathrm{d}E\,.\nonumber
\end{eqnarray}
The resulting expression for the net current is%
\begin{equation}
I=\frac{e}{4\pi\hslash}\sum_{n_{0},n}\int\biggl\{
\left(  \Sigma_{n_{0}%
,n,L}^{<}-\Sigma_{n_{0},n,R}^{<}\right)_{\alpha}
\left[G_{n,n_{0}}^{r}%
(E)-G_{n,n_{0}}^{a}(E)\right]
\biggr\}dE, \label{31}%
\end{equation}
\end{widetext}
which is derived from equation~(\ref{30}) using the same damping factor for each
lead ($\Gamma_{L}=\Gamma_{R}=\Gamma$
).

For the present case of zero temperature the lesser self--energy may be recast
in terms of the Heaviside step function $\theta(x)$ as%
\begin{equation}
\Sigma^{<}_{n_0,n_0,\alpha}
=\mathrm{i}\Gamma_{\alpha}\theta\left(  \epsilon_{\mathrm{F}\alpha
}-\varepsilon\right)  \delta_{n_{0},0}\,, \label{32}%
\end{equation}
where $\epsilon_{\mathrm{F}\alpha}$ is the Fermi energy on lead $\alpha$ and
the Kronecker delta, $\delta_{n_{0},0}$, signifies that the nanomechanical
oscillator is initially in its ground state, $n_{0}=0$.

Inserting this equation in the above equation, the result is%

\begin{eqnarray}
I&=&\frac{e}{2\pi\hslash}\int  dE \left[T_{0}(E)\right]\nonumber\\
&&\times \left[
\theta\left(  \epsilon_{\mathrm{F}L}%
+{\mathchoice{{\textstyle{\frac12}}}{{\textstyle{\frac12}}}{{\scriptstyle{1/2}}}{{\scriptscriptstyle{1/2}}}}%
\hslash\omega-E\right)
-\theta\left(  \epsilon_{\mathrm{F}R}%
+{\mathchoice{{\textstyle{\frac12}}}{{\textstyle{\frac12}}}{{\scriptstyle{1/2}}}{{\scriptscriptstyle{1/2}}}}%
\hslash\omega-E\right)
\right]\quad \label{33}%
\end{eqnarray}
where
\[
T_{0}(E)=\frac{i\Gamma}{2}(G_{0,0}^{r}(E)-G_{0,0}^{a}(E))=\Gamma\lbrack
G_{0,0}^{r}(E)\Gamma G_{0,0}^{a}(E)]
\]

Using the expression for the retarded and advanced Green's function, the
expression for the net current becomes%
\begin{eqnarray}
\lefteqn{I=\frac{e}{2\pi\hslash}\sum_{n}\left\vert \Phi_{0,n}\right\vert ^{2}}\nonumber\\%
&\times&\int_{\epsilon_{\mathrm{F}R}+\frac{1}{2}\hslash\omega}^{\epsilon_{\mathrm{F}L}+\frac{1}{2}\hslash\omega}
\frac{\Gamma^{2}dE}{\left[  E-\epsilon_{0}-\left(
n+\frac{1}{2}\right)  \hslash\omega+\Delta\right]  ^{2}+\Gamma^{2}}
\label{34}%
\end{eqnarray}
After performing the integral in the above expression, the final result is
written as%
\begin{eqnarray}
I&=&\frac{e\Gamma}{2\pi\hslash}\sum_{n}\left\vert \Phi_{0,n}\right\vert
^{2}\Biggl[  \tan^{-1}\left(  \frac{\epsilon_{\mathrm{F}L}-\epsilon
_{0}-n\hslash\omega+\Delta}{\Gamma}\right)\Biggr.\nonumber\\  
&&\hspace{40pt}-\Biggl.\tan^{-1}\left(  \frac
{\epsilon_{\mathrm{F}R}-\epsilon_{0}-n\hslash\omega+\Delta}{\Gamma}\right)
\Biggr]  \label{35}%
\end{eqnarray}
The model represent the interplay between two physical time scales of the
system, the oscillator frequency and the tunneling rate. This model also shows
the very interesting interplay between two physical length scales of the
system, zero point amplitude and zero point displacement, which actually
affected by the weak and strong coupling dynamics.

The zero frequency shot-noise has been derived and applied successfully in
many examples of the transport dynamics of nanoscopic systems\cite{17,24,26}.
This is given as%
\begin{widetext}
\begin{equation}
S(0)  = \frac{e^{2}}{\pi\hslash}%
{\displaystyle\int}
\biggl\{\left[f_{L}(\varepsilon)(1-f_{L}(\varepsilon))+f_{R}(\varepsilon)(1-f_{R}%
(\varepsilon))\right]T_{0}(E)
+\left[f_{L}(\varepsilon)-f_{R}(\varepsilon)\right]^{2}%
\left(1-T_{0}(E)\right)T_{0}%
(E)\biggr\}\,dE\,,\label{36}
\end{equation}
Using the values of transmission coefficients at zero temperature, the above
expression can be simplified as%
\begin{eqnarray}
S(0) &=& \frac{e^{2}}{\pi\hslash}%
{\displaystyle\int_{\epsilon_{\mathrm{F}R}+\frac{1}{2}\hslash\omega
}^{\varepsilon_{F_{L}}+\frac{1}{2}\hslash\omega}}
(1-T_{0}(E))T_{0}(E)\,dE\label{37}\\
&=& \frac{e^{2}}{\pi\hslash}\int_{\epsilon_{\mathrm{F}R}+\frac{1}{2}%
\hslash\omega}^{\epsilon_{\mathrm{F}L}+\frac{1}{2}\hslash\omega}
\left\{
\sum_{n}\frac{\left\vert \Phi_{0,n}\right\vert ^{2}\Gamma^{2}}{\left[
E-\epsilon_{0}-\left(  n+\frac{1}{2}\right)  \hslash\omega+\Delta\right]
^{2}+\Gamma^{2}}
-\left(  \sum_{n}\frac{\left\vert \Phi_{0,n}\right\vert ^{2}\Gamma^{2}%
}{\left[  E-\epsilon_{0}-\left(  n+\frac{1}{2}\right)  \hslash\omega
+\Delta\right]  ^{2}+\Gamma^{2}}\right)  ^{2}%
\right\}\,dE  \nonumber
\end{eqnarray}
After integrating the above expression, we arrive at the final result%
\begin{equation}
S(0)=\frac{e^{2}\Gamma}{\pi\hslash}\left\{  \sum_{n}S_{n}(0)+\sum_{n>m}%
S_{n,m}(0)\right\}  , \label{38}%
\end{equation}
where $S_{n}(0)$ and $S_{n,m}(0)$ are defined as%

\begin{eqnarray}
S_{n}(0)&=&\left\vert \Phi_{0,n}\right\vert ^{2}\Biggl\{\Biggl[
\tan^{-1}\left(  \frac{\epsilon_{\mathrm{F}L}-\epsilon_{0}-n\hslash
\omega+\Delta}{\Gamma}\right)\Biggr.  
-\tan^{-1}\left(  \frac{\epsilon_{\mathrm{F}%
R}-\epsilon_{0}-n\hslash\omega+\Delta}{\Gamma}\right)\Biggr]\nonumber \\%
&&\hspace{40pt}+\frac{\left\vert \Phi_{0,n}\right\vert ^{2}}{2}\Biggl[  \frac{\Gamma
(\epsilon_{\mathrm{F}L}-\epsilon_{0}-n\hslash\omega+\Delta)}{(\epsilon
_{\mathrm{F}L}-\epsilon_{0}-n\hslash\omega+\Delta)^{2}+\Gamma^{2}}
+\tan^{-1}\left(\frac{\epsilon_{\mathrm{F}L}-\epsilon_{0}-n\hslash\omega+\Delta}{\Gamma
}\right)\Biggr]\nonumber \\
&&\hspace{40pt}\Biggl.-\frac{\left\vert \Phi_{0,n}\right\vert ^{2}}{2}\Biggl[  \frac{\Gamma
(\epsilon_{\mathrm{F}R}-\epsilon_{0}-n\hslash\omega+\Delta)}{(\epsilon
_{\mathrm{F}R}-\epsilon_{0}-n\hslash\omega+\Delta)^{2}+\Gamma^{2}}
-\tan^{-1}\left(\frac{\epsilon_{\mathrm{F}R}-\epsilon_{0}-n\hslash\omega+\Delta}{\Gamma
}\right)\Biggr]
\Biggr\}\nonumber
\end{eqnarray}
\begin{eqnarray}
S_{n,m}(0)&=&
\frac{\left\vert \Phi_{0,n}\right\vert ^{2}\left\vert \Phi_{0,m}\right\vert
^{2}\Gamma^{2}}{i((n-m)\hslash\omega)((n-m)\hslash\omega+2i\Gamma)}
\Biggl\{
\frac{1}{2}\ln\left[  \frac{(\epsilon_{\mathrm{F}L}-\epsilon_{0}%
-n\hslash\omega+\Delta)^{2}+\Gamma^{2}}{(\epsilon_{\mathrm{F}R}-\epsilon
_{0}-n\hslash\omega+\Delta)^{2}+\Gamma^{2}}\right]\Biggr.\nonumber\\
&&\hspace{60pt}\Biggl.+i\left(\tan^{-1}\left[\frac{\epsilon_{\mathrm{F}L}-\epsilon_{0}-n\hslash\omega+\Delta
}{\Gamma}\right]
-\tan^{-1}\left[\frac{\epsilon_{\mathrm{F}R}-\epsilon_{0}-n\hslash
\omega+\Delta}{\Gamma}\right]\right)
\Biggr\}\nonumber \\
&&+\frac{\left\vert \Phi_{0,n}\right\vert ^{2}\left\vert \Phi_{0,m}\right\vert
^{2}\Gamma^{2}}{i((m-n)\hslash\omega)((m-n)\hslash\omega+2i\Gamma)}
\Biggl\{
\frac{1}{2}\ln\left[  \frac{(\epsilon_{\mathrm{F}L}-\epsilon_{0}%
-m\hslash\omega+\Delta)^{2}+\Gamma^{2}}{(\epsilon_{\mathrm{F}R}-\epsilon
_{0}-m\hslash\omega+\Delta)^{2}+\Gamma^{2}}\right]\Biggr.\nonumber \\
&&\hspace{60pt}\Biggl.+i\left(\tan^{-1}\left[\frac{\epsilon_{\mathrm{F}L}-\epsilon_{0}-m\hslash\omega+\Delta
}{\Gamma}\right]
-\tan^{-1}\left[\frac{\epsilon_{\mathrm{F}R}-\epsilon_{0}-m\hslash
\omega+\Delta}{\Gamma}\right]
\right)
\Biggr\}\nonumber
\end{eqnarray}
\end{widetext}
The corresponding Fano factor can be calculated from the zero frequency noise
and the net current flowing through the system, which is defined as%

\begin{equation}
F=\frac{S(0)}{2eI} \label{39}%
\end{equation}

\section{Discussion of results}

The current-voltage (I-V) characteristics, shot noise spectrum and the
corresponding Fano factor of a resonant tunnel junction coupled to a
nanomechanical oscillator are shown graphically for different values of
coupling strength, using the same parameters as: the single energy level of
the dot $\epsilon_{0}=0.5$, the characteristic frequency of the oscillator
$\hslash\omega=0.1$, the damping factor $\Gamma=0.1\hslash\omega$ and the
chemical potentials $0\leq\epsilon_{\mathrm{F}L}\leq1$ and $\epsilon
_{\mathrm{F}R}=0$. These are chosen to illustrate the physics of such systems
rather than to represent a specific implementation. The oscillator induced
resonance effects are clearly visible in the numerical results. It must be
noted that we have obtained these results in the regime of strong and zero or
weak coupling of the oscillator with the electrons on the dot. The coupling
between the leads and the dot is considered to be symmetric and we assume that
the electrons in the leads are at zero temperature and have constant density
of states. With increasing coupling strength, the number of additional steps
also increases while for zero or weak coupling we find only the basic
resonance step. This confirms the effect of the coupling between the electrons
on the dot and the single oscillator mode where higher energy electrons are
able to drop to the dot energy by creation of phonons.

Closer analytical examination of the expression for the current and shot noise
(35 \& 38) shows that the main resonance steps occur when the applied voltage,
$\epsilon_{FL}$ is equal to the energy eigenvalues of the coupled dot electron
and the nanomechanical oscillator. The main step $(n=0)$ is given by the
Lorentzian form with its center at the $\epsilon_{FL}=\epsilon_{0}-\Delta$,
known as a Breit-Wigner resonance. The additional steps due to the emission of
phonons can be seen on the positive energy side with $\epsilon_{FL}%
=\epsilon_{0}-\Delta+n\hslash\omega$ where $\omega$ is the characteristic
frequency of the nanomechanical oscillator.
\begin{figure}[htb]
\includegraphics[width=\columnwidth]{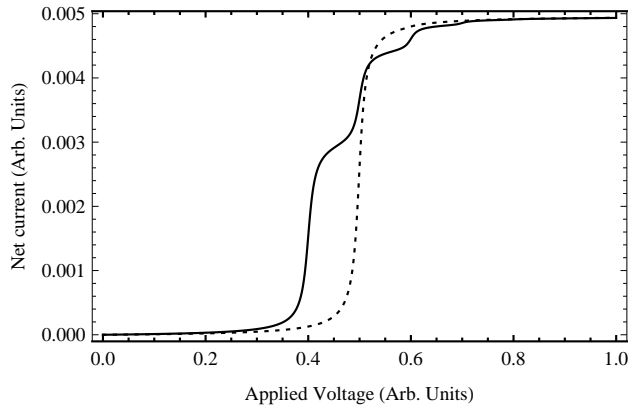}
\caption{\label{fig1} Current-voltage characteristics (dimensionless) as a function of
applied voltage $\epsilon_{FL}$ (in arbitrary units), coupling strength
$\eta=0.1\hslash\omega$(dotted line) and 1$\hslash\omega$ (solid line). Gate
voltage $\epsilon_{0}=0.5,$ oscillator energy $\hslash\omega=0.1$ and
self-energy $\Gamma=0.1\hslash\omega.$
}
\end{figure}
We have shown the I-V characteristics of the NEMS device against applied bias
for different values of the coupling strength in Fig.~\ref{fig1}. The main resonance
step is the elastic or zero phonon transition. The amplitude of the additional
steps is much smaller than the basic resonance step. The electrons that tunnel
onto the dot can only excite the oscillator mode as at zero temperature there
are no phonons available to be absorbed. Moreover, we have seen that with
increasing coupling strength, the number and intensity of the additional steps
increases but their intensity always remains much smaller than the main step.
The steps in the current characteristics vanish if the upper electrochemical
potential is smaller than the dot energy plus the oscillator energy.
\begin{figure}[htb]
\includegraphics[width=\columnwidth]{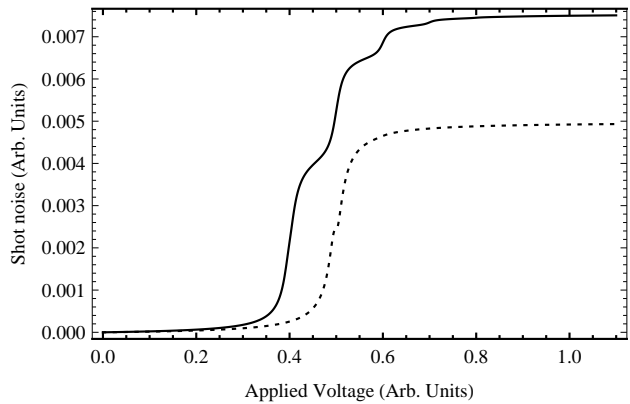}
\caption{\label{fig2}Shot noise (dimensionless) as a function of applied voltage
$\epsilon_{FL}$ (in arbitrary units), with gate voltage $\epsilon_{0}$=0.5,
oscillator energy $\hslash\omega=0.1,$ self-energy $\Gamma=0.1\hslash\omega$,
and coupling strength $\eta=0.1\hslash\omega$ (dotted line) and
1$\hslash\omega$ (solid line).}
\end{figure}

\begin{figure}[htb]
\includegraphics[width=\columnwidth]{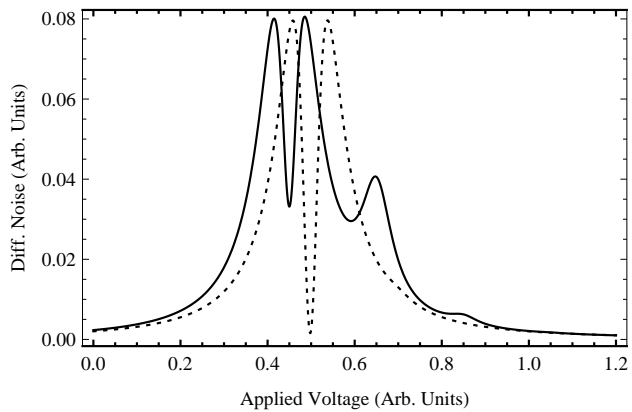}
\caption{\label{fig3}Differential shot noise (dimensionless) as a function of applied
voltage $\epsilon_{FL}$ (in arbitrary units) with gate voltage $\epsilon
_{0}=0.5,$ oscillator energy $\hslash\omega=0.2,$ self-energy $\Gamma
=0.2\hslash\omega$ and coupling strength $\eta=0.1\hslash\omega$ (dotted
line) and $\eta=0.5\hslash\omega$ (solid line).}
\end{figure}

Next, we have shown the shot noise as a function of applied bias in Fig.~\ref{fig2},
which exhibits the single step in the presence of weak or zero
electron-oscillator coupling while the number of additional steps increases
with increasing coupling strength. Obviously, the shape of the shot noise
curve is similar to that of the net current, as shown in Fig.~\ref{fig1}. The only
difference is associated with their behavior above the resonance point, where
the noise power for the strong coupling case can exceed the shot noise for the
zero phonon case. We also show the differential shot noise against applied
bias for different values of the coupling strength(in Fig.~\ref{fig3}). In the absence
of phonons, when the transport is coherent, the shot noise spectrum exhibits
two peaks separated by an antiresonance and located symmetrically around the
position where the current step associated with the single level of the dot is
located. The origin of such an antiresonance in the noise spectrum is
associated with the fact that no noise is generated when the transmission via
the dot state is perfect, T = 1 or zero. Due to the presence of the
nanomechanical oscillator, the main resonance peaks are shifted by $\Delta.$
Meanwhile, the sharp peaks are now accompanied by a set of additional peaks.
We note that the separation between the differential shot noise peaks is set
by the frequency of the oscillator. This phenomenon can be explained on the
basis of phonon emission during electron-phonon coupling, indicating that a
new channel has opened and contributes to the resonant tunneling process.

\begin{figure}[htb]
\includegraphics[width=\columnwidth]{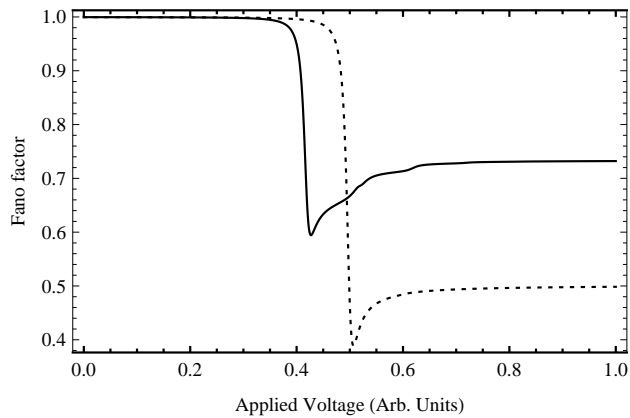}
\caption{\label{fig4}Fano factor as a function of applied voltage $\epsilon_{FL}$ (in
arbitrary units) with gate voltage $\epsilon_{0}$=0.5, oscillator energy
$\hslash\omega=0.1,$ self-energy $\Gamma=0.1\hslash\omega$ and coupling
strength $\eta=0.1\hslash\omega$ (dotted line) and $\eta=1\hslash\omega$
(solid line).
}
\end{figure}

Finally, we have shown the Fano factor against applied bias for different
values of the coupling strength. Information about the statistical properties
of the electrons is included in the Fano factor, which is plotted in 
Fig.~\ref{fig4}.
Since in our model all the interactions between the current carriers are
neglected, such electron correlations are associated only with the Coulomb
blockade. This principle is related to the fact that one electron feels the
presence of the others, since it cannot occupy the state on the dot already
occupied by the electron. The crossover in the shot noise power from
Poissonian (F =1) to sub-Poissonian (F%
$<$%
1) is always observed after the first step in the current voltage dependence.
This implies that electrons tunnel in a correlated way in the NEMS\ device.
The most important result is the significant enhancement of the Fano factor
due to the phonon effects, observed for $\epsilon_{FL}>\epsilon_{0}%
-\Delta+n\hslash\omega$, where the multi-channel process reduces electron
correlations compared with the single channel case. Moreover, the shift
$\Delta$ can also be easily recognized in Fig.~\ref{4}.

\section{Summary}

In this work, we analyzed the current noise characteristics of a resonant
tunnel junction coupled to a nanomechanical oscillator by using the
non-equilibrium Green's function approach without treating the electron-phonon
coupling as a perturbation. We have derived an analytical expressions for the
net current flowing through the system and for the shot noise. This enables us
to see the effects of the coupling of the electrons to the oscillator on the
dot and the tunneling rate of electrons. We show the numerical results for
very weak and strong coupling strength. We have found additional steps or
peaks due to coupling of single phonon mode which are absent for very weak or
no coupling strength. We also discuss the corresponding Fano factor as a
function of applied bias which shows thermal or poissonian behavior to
non-thermal or sub-poissonian behavior.

\end{document}